\documentclass{ws-rv9x6}
\usepackage{ws-rv-van} % numbered citation/references
\makeindex
\begin{document}
\chapter[The third moment of current fluctuations in a tunnel junction]{The third moment of current fluctuations in a tunnel junction: experiments in the classical and quantum regimes}

\author{B. Reulet, J. Gabelli}
\address{Laboratoire de Physique des Solides, CNRS UMR8502,\\
 Universit\'e de Paris-Sud 11, Orsay, France}
\author[B. Reulet, J. Gabelli, L. Spietz and D.E. Prober]{L. Spietz\footnote{Present address: NIST, Boulder CO, USA.} and D.E. Prober}
\address{Department of Applied Physics, Yale University, New Haven CT, USA}

\begin{abstract}
We report the first experimental data of the third moment of current fluctuations in a tunnel junction. We show that both in the classical and quantum regimes (low or high frequency as compared to voltage), it is given by $S_{I^3}=e^2I$. We discuss environmental effects in both regimes.
\end{abstract}

\body

\section{Introduction}

In a seminal article \cite{LL}, Levitov and Lesovik asked themselves the following, simple question: what is the probability that $n$ electrons cross a coherent conductor in a time $\tau$ ? The final answer came three years later, by a quantum mechanical treatment of the electronic wavefunctions, of the scattering by the sample and even of the charge detector \cite{Levitov}. Despite a fully quantum approach, the result of the calculation was quite disappointing: electrons behave as classical particles. At zero temperature, electrons try to enter the device at a pace $eV/h$, and for each attempt they succeed with a probability $p=|t|^2$, where $t$ is the transmission probability of the electronic wave through the scattering region ($e$ is the electron charge and $h$ Planck's constant). The distribution is binomial, $P(n,\tau)=C_N^np^n(1-p)^n$ where $N=\tau eV/h$ is the total number of attempts and $C_N^n=N!/[n!(N-n)!]$. Only the long time limit ($N\gg1$) was considered. The result of ref. \cite{Levitov} implies non-Gaussian current fluctuations in the sample, the simplest measure of this non-Gaussian aspect being given by the existence of a third moment of the transferred number of particles: $\langle \delta n^3\rangle=p(1-p)(1-2p)N$, with $\delta n=n-\langle n\rangle$. This reflects the asymmetry of the fluctuations around the mean value: the distribution is skewed, one direction is preferred. In the following we will only consider tunnel junctions, which are the simplest system that exhibit shot noise, and for which the second moment has been thoroughly studied \cite{Lafe}.

\section{The third moment in the classical regime}

\begin{figure}
\centerline{\psfig{file=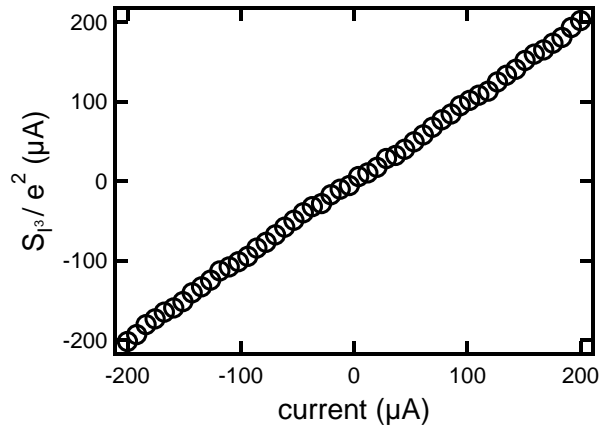,width=8cm}}
\caption{Normalized third moment of current fluctuations $S_{I^3}$ vs. dc current measured at temperature $T=4.2$ K and low frequency ($0.01 - 1$ GHz). It is obtained from the top panel of Fig. 2 after subtraction of the environmental contributions, see Eq. (\ref{eq_env}).}
\label{S3lf}
\end{figure}

For a tunnel junction one has $p\ll1$ and the probability distribution $P(n,\tau)$ is Poissonian, with $\langle \delta n^3\rangle=\langle \delta n^2\rangle=\langle n\rangle$. This translates into current fluctuations having a third moment spectral density $S_{I^3}=e^2I$ where $I=\langle I\rangle=\langle n\rangle e/\tau$ is the dc current (the spectral density of the variance is $S_{I^2}=eI$). The relation between the fluctuations of the current integrated in the bandwidth from $f_1$ to $f_2$ and the spectral density is non-trivial: $\langle\delta I^2\rangle=2(f_2-f_1)S_{I^2}$ (the factor $2$ come from positive and negative frequencies) and $\langle\delta I^3\rangle=3B^2S_{I^3}$ with $B=f_2-2f_1$ if $f_2>2f_1$ and $B=0$ otherwise (we have confirmed this unusual dependence of the measurement on the frequency cutoffs\cite{Reulet}). As a consequence, the relation between the measured current distribution and the real one is very subtle\cite{avalanches}. We have experimentally confirmed the prediction $S_{I^3}=e^2I$, see \cite{Reulet}. We show on Fig. \ref{S3lf} the result for a tunnel junction measured in the bandwidth 0.01-1 GHz at temperature $T=4.2$ K. At equilibrium ($I=0$), positive and negative current fluctuations are equiprobable, so $S_{I^3}^{eq}=0$. Subsequent work has confirmed our result \cite{Reznikov}.

\begin{figure}
\centerline{\psfig{file=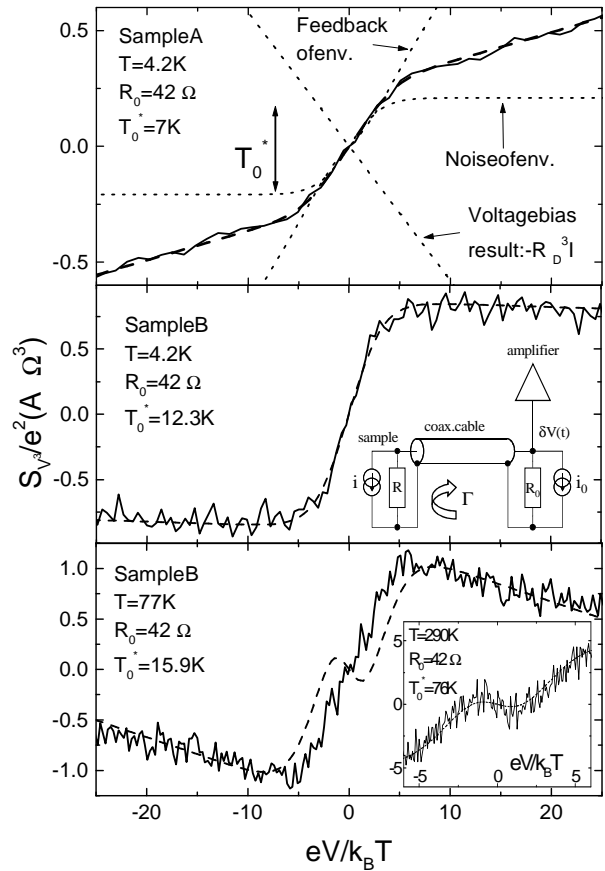,width=8cm}}
\caption{Measurement of $S_{V^3}(eV/k_BT)$ (solid lines). The dashed lines corresponds to the best fit with Eq. (\ref{eq_env}). The dotted lines in the top plot correspond to the different contributions to $S_{V^3}$ (see text). Inset of the middle plot: schematics of the measurement circuit. The effective noise temperature of the environment $T_0^*$ is equal to the real noise temperature times the reflection coefficient $\Gamma$ of the sample when the cable between the sample and the amplifier is long (which was not the case at room temperature).}
\label{S3T}
\end{figure}

Finite temperature affects the transport, so the probability for charge transfer is temperature dependent. In a tunnel junction, the spectral density of the variance (second moment) of current fluctuations is given by: $S_{I^2}=GeV\coth eV/(2k_BT)$, with $T$ the temperature and $G$ the sample's conductance. Noise vs. voltage evolves continuously from the equilibrium, Johnson-Nyquist noise $S_{I^2}^{eq}=2k_BTG$ at $V=0$, to the full shot noise $S_{I^2}^{shot}=eI$ at large voltage, $V\gg k_BT/e$ (for a review on $S_{I^2}$, see \cite{Blanter}). Strikingly, $S_{I^3}$ has a very different behavior: it is totally temperature independent, i.e. thermal fluctuations of current are  symmetric. Thus $S_{I^3}$ can be measured at high temperature and/or low voltage to reveal the charge of the carriers, contrary to $S_{I^2}$ which requires $eV>k_BT$. We have measured $S_{I^3}$ from 300 K down to 50 mK and indeed found no temperature dependence \cite{Reulet,Bertrand_Houches}. We show on Fig. \ref{S3T} measurements at 300 K, 77 K and 4.2 K on different samples. They all indicate
$S_{I^3}=e^2I$ independent of temperature, despite the environmental contributions which do depend on temperature (see below).

\section{Environmental effects in the classical regime}

Usual theories consider the fluctuations of charge transfer at a fixed bias voltage. This rarely describes an experimental situation: voltage or current sources have a finite output impedance and voltage/current amplifiers a finite input impedance, in particular at high frequency. This usually does not matter for average quantities or variances: the average current under voltage bias is $\langle I\rangle_V=GV$, the average voltage under current bias is $\langle V\rangle_I=RI$, and the conductance $G$ and the resistance $R$ are inverse of each other: $R=1/G$. Similarly, voltage fluctuations under current bias (i.e., sample is open at ac) and current fluctuations under voltage bias (i.e., sample is short-circuited at ac) are proportional, $\langle\delta I^2\rangle_V=G^2\langle\delta V^2\rangle_I$, or in terms of spectral densities, $S_{I^2}=G^2S_{V^2}$. Finite source or amplifier impedances lead to a simple rescaling by a resistance ratio, and the noise of the source or amplifier adds incoherently to that of the sample: $S_{V^2}=R_D^2(S_{I^2}+S_{i_0^2})$ where $R_D=RR_0/(R+R_0)$ is the parallel combination of the sample's and the environmental resistances and $S_{i_0^2}$ is the noise spectral density of the amplifier, for the circuit depicted in the middle inset of Fig. \ref{S3T}.

\begin{figure}
\centerline{\psfig{file=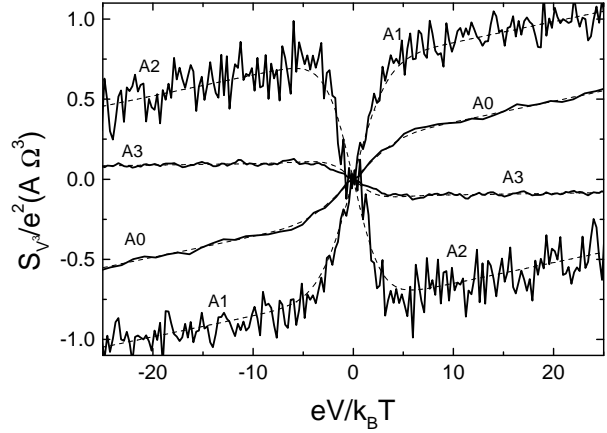,width=8cm}}
\caption{Measurement of $S_{V^3}(eV/k_BT)$ for sample A at T=4.2 K (solid lines). A0: no ac excitation (same as Fig. 1, top panel). A1: with an ac excitation at frequency $\Omega/2\pi$ such that $\cos2\Omega\Delta t=+1$; A2: $\cos2\Omega\Delta t=-1$. The ac excitation at frequency $\Omega$ simulates the noise of the amplifier sent to the sample, and demonstrates the influence of the environmental noise. The frequency dependence with the delay time $\Delta t$ explains why $T_0$ has to be replaced by $T_0^*$ in the environmental contribution. A3: no ac excitation but a $63\;\Omega$ resistor in parallel with the sample. This demonstrates the effect of the environmental impedance. The dashed lines corresponds to fits with Eq. (\ref{eq_env}). A1 and A2 differ from A0 only by a term $\propto dS_{I^2}/dV$. A3 is deduced from A0 with no fitting parameter.}
\label{S3env}
\end{figure}

For the third and higher order moments, the story becomes more complicated: in the presence of a finite external impedance, the voltage across the sample fluctuates, which modifies the probability of charge transfer \cite{Nazarov,Beenakker}. There are two contributions to the voltage fluctuations: noise of the environment (Johnson noise of the external impedance or current noise of the amplifier, $i_0$), and current fluctuations of the sample itself that are converted into voltage fluctuations by the external impedance $R_0$. The first environmental contribution to $S_{I^3}$ expresses how the amount of noise generated by the sample (the variance $S_{I^2}$) is modulated by a small fluctuating voltage $R_Di_0$, and is $\propto R_DS_{i_0^2} dS_{I^2}/dV$. It disappears if the environment is perfectly quiet ($i_0=0$), but exists even if the external impedance is zero. In contrast in the second contribution, the feedback of the environment, the noise source is the sample itself, so it does not depend on the environmental noise temperature, but the environmental impedance converts the sample current fluctuations into voltage fluctuations, resulting in a contribution to $S_{I^3}$ that is $\propto R_DS_{I^2}dS_{I^2}/dV$. As a result, the spectral density of the third moment of voltage fluctuations measured across the resistance $R_0$ in parallel with the sample is given by:

\begin{equation}
S_{V^3}=R_D^3\left[-S_{I^3}+ 3R_DS_{i_0^2}\frac{dS_{I^2}}{dV}+ 3R_DS_{I^2}\frac{dS_{I^2}}{dV}\right]
\label{eq_env}
\end{equation}
We have experimentally demonstrated the existence of these two mechanisms and their contribution to $S_{V^3}$, see Fig. \ref{S3env}. Note that the absence of environmental corrections to the average current and to $S_{V^2}$ are only approximate. The feedback mechanism indeed leads to an environmental correction to the dc current, better known as dynamical Coulomb blockade \cite{Ingold, Bertrand_Houches, GR2}. Dynamical Coulomb blockade is hardly visible for low impedance samples but is important for the third moment. Environmental corrections to the second moment have not yet been observed\cite{Lafe}.

\section{The third moment in the quantum regime}

\begin{figure}
\centerline{\psfig{file=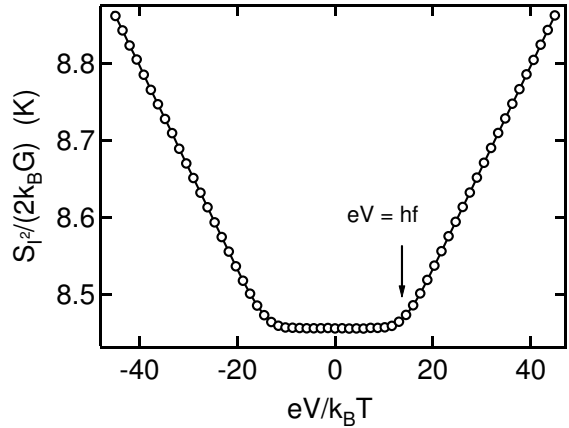,width=8cm}}
\caption{Measurement of $S_{I^2}(V)/(2k_BG)$ at $f=5.5-6.5$ GHz and $T=20$ mK. The rounding of the curve is not limited by the temperature but by the bandwidth of the measurement (1 GHz is 50 mK). The value at $V=0$ is $T_N+hf/(2k_B)=T_N+0.15$K with $T_N$ the amplifier's noise temperature}
\label{S2hf}
\end{figure}

The result of ref.\cite{Levitov} that electrons behave as marbles is valid for $\tau\rightarrow +\infty$, i.e. in the low frequency limit $f\ll eV/h$. At high frequency, one may hope that quantum mechanics will dictate the statistics of charge transfers. As a matter of fact, the spectral density of the variance of current fluctuations at finite frequency $f$ is given by $S_{I^2}(V,f)=[S_{I^2}(V+hf/e,0)+S_{I^2}(V-hf/e,0)]/2$ for a tunnel junction. At zero temperature, it is voltage-independent and equal to $Ghf$ as long as $|eV|<hf$, since electrons of energy $eV$ cannot emit photons of greater energy $hf>eV$. The low voltage plateau on $S_{I^2}$ is clearly of quantum nature. We show on Fig. \ref{S2hf} the observation of the plateau, in a way clearer than previously reported data \cite{Schoelkopf,Portier,GR1} thanks to the high $hf/(k_BT)=15$ ratio that we have been able to achieve ($f=6$ GHz, $T=20$ mK).

$S_{I^2}$ at finite frequency results from the product of the current at frequency $f$ by the current at frequency $-f$. In classical mechanics, those two are c-numbers; in quantum mechanics they are operators. These operators do not commute, since the current operators taken at two different times do not commute. Thus, the correlators $S_+(f)=\langle\hat I(f)\hat I(-f)\rangle$, $S_-(f)=\langle\hat I(-f)\hat I(f)\rangle$, or more generally $S_\alpha(f)=\alpha S_+ +(1-\alpha)S_-$ are all different (but are identical in the classical limit), and one has to face the problem of knowing which experimental setup measures which combination. However, for simple systems $S_+$ and $S_-$ are related by $S_+-S_-=Ghf$, so all the correlators differ only by a constant $\propto Ghf$, independent of both temperature and voltage, thus indistinguishable in many experiments (although some setups correspond to a well established value of $\alpha$ \cite{Deblock,Gavish,Lesovik}). This simplification does not occur for higher order cumulants, and the various ways of ordering the operators leads to different voltage dependance \cite{Zaikin1,Zaikin2,Hekking}.

\begin{figure}
\centerline{\psfig{file=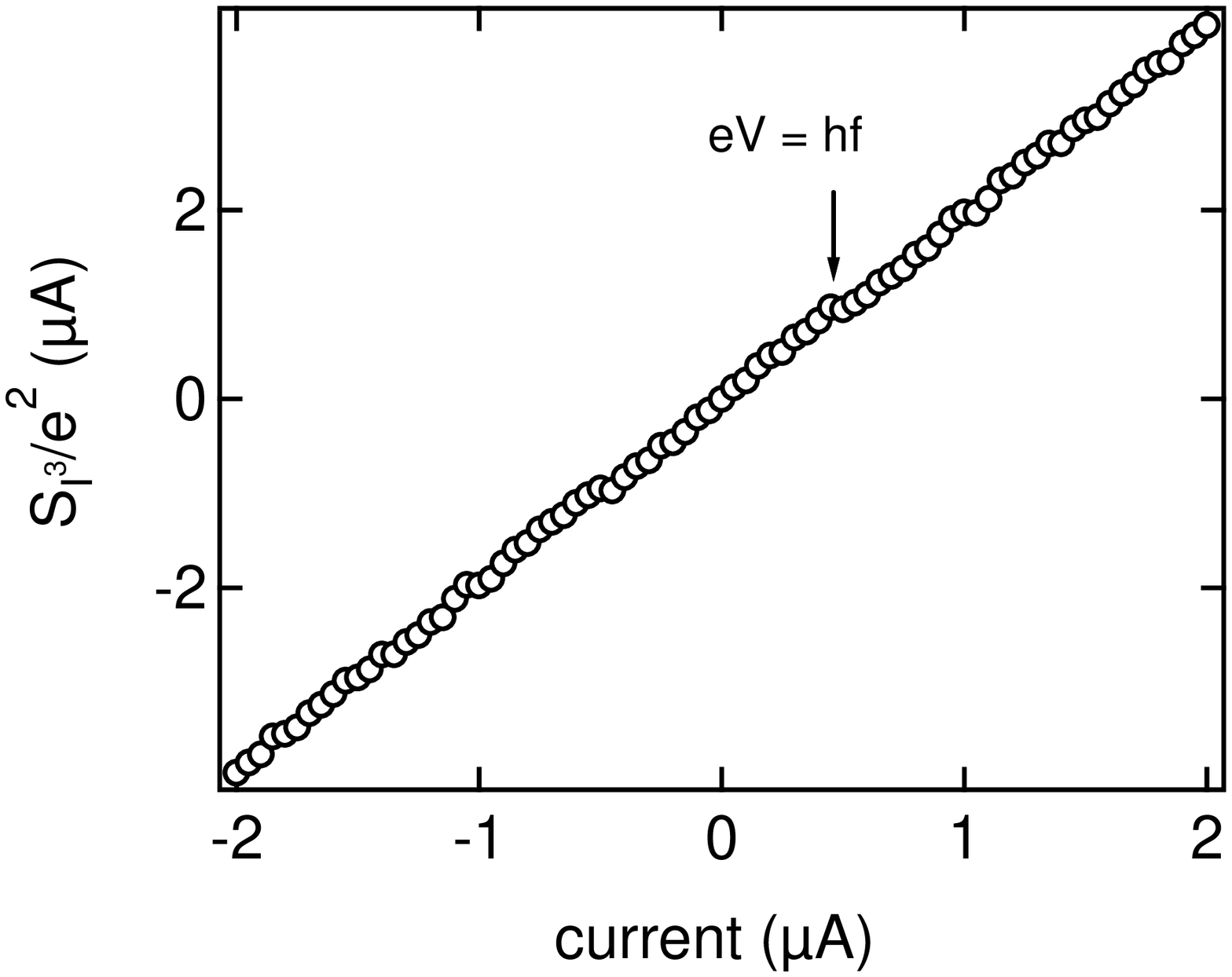,width=8cm}}
\caption{Normalized third moment of current fluctuations $S_{I^3}(0,f)$ measured at very low temperature $T=27$ mK and high frequency $f=5.5 - 6.5$ GHz.}
\label{S3hf}
\end{figure}

We have performed a measurement of $\langle\delta V(\epsilon)\delta V(f-\epsilon)\delta V(-f)\rangle$ ($\epsilon < k_BT/h \ll f$) using linear amplifiers and square law detectors, from which we deduce $S_{I^3}(\epsilon\rightarrow 0,f)$. We observe that $S_{I^3}(0,f)=e^2I$ regardless of the frequency $f$, see Fig. \ref{S3hf}. It is very surprising that $S_{I^3}$ is finite even though $eV<hf$. This means that $S_{I^3}$ can be non-zero even if no photon of frequency $f$ is emitted by the sample. A non-zero $S_{I^3}(0,f)$ implies the existence of correlations between the low frequency current fluctuations $I(\epsilon\rightarrow 0)$ and  the  high frequency noise $I(f)I(-f)$ i.e., for $eV<hf$, between the low frequency current fluctuations and the high frequency zero point motion of electrons. This phenomenon has never been discussed. Our measurement directly raises the problem of how to relate an experimental setup to a quantum mechanical prediction. Had we used a photo-detector (which clicks each time it absorbs a photon) instead of a linear amplifier followed by a square law detector, we would have obtained $S_{I^3}(0,f<eV/h)=0$ since no photon would have been detected for $eV<hf$ \cite{upon}.

\section{Environmental effects in the quantum regime: the noise susceptibility}

Since we have seen that environmental effects are crucial to $S_{I^3}$ measured at low frequency, it is necessary to investigate what they become in the quantum regime. The environmental contributions to $S_{I^3}$ involve the modulation of $S_{I^2}$ by a time-dependent voltage $v_0$. This modulation is given by $\delta S_{I^2}=v_0\chi$ where $\chi$, the noise susceptibility, describes by how much the noise is modulated by the ac excitation. In the adiabatic limit, i.e. provided the noise ``follows" the excitation, $\chi=dS_{I^2}/dV$. This happens as long as the frequency $f_0$ of the ac excitation is smaller than some characteristic frequency $f_c$.

In order to have an intuition of $f_c$, let us first consider the simple case of a macroscopic sample, for which the noise is the equilibrium value $S_{I^2}^{eq}=2k_BTG$,  biased by a voltage $V+v_0\cos2\pi f_0 t$. The oscillating voltage creates oscillating Joule heating of the sample, thus oscillating temperature and oscillating noise (measured at frequency $f\gg f_0$). In this case $f_c$ is clearly the inverse of the thermalization time. The linear response of the sample noise temperature to the ac Joule heating defines the noise thermal impedance between the system and its thermal environment, and allows for a direct measurement of thermalization times, e.g. by electron-phonon interaction or by diffusion of hot electrons \cite{NTI}. This also explains the existence of an intrinsic $S_{I^3}$ in a diffusive wire or a chaotic cavity \cite{Pilgram, Nagaev}, where the variation of the Joule heating is given by the current fluctuations in the sample times the bias voltage, and the cutoff $f_c$, proportional to the inverse diffusion time, determines the frequency dependence of $S_{I^3}$.

\begin{figure}
\centerline{\psfig{file=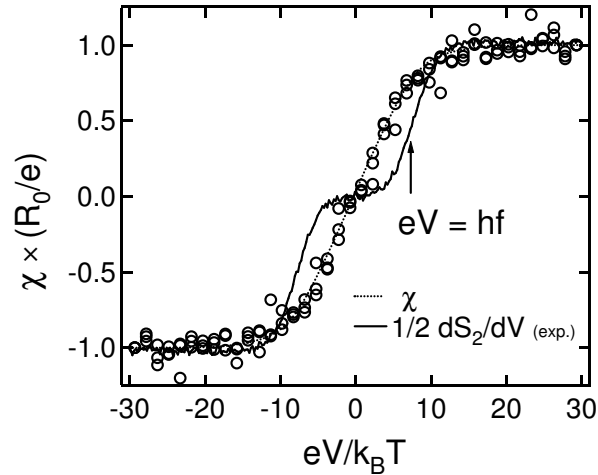,width=8cm}}
\caption{Normalized noise susceptibility $\chi$ vs. normalized dc bias. Voltage fluctuations are detected in the range $f=5.5-6.5$ GHz and $|f_0-f|=0.01-0.5$ GHz, whereas rf excitation is of amplitude $V_{ac}$ at frequency $f_0=6$ GHz. Symbols: data for various levels of ac excitation ($eV_{ac}/(hf_0) = 0.85$,
$0.6$ and $0.42$). Dotted lines: theoretical prediction for $\chi$. Solid line: measured $(1/2)dS_2(f)/dV$, as a comparison.}
\label{chi}
\end{figure}

This clear picture however does not apply in the quantum regime, for which $f\gg f_0$ is not fulfilled. In that case, we have shown that $\delta S_{I^2}$ is given by a correlator involving current taken at two different frequencies separated by $f_0$: $\delta S_{I^2}=\langle \hat I(f)\hat I(f_0-f)\rangle+\langle \hat I(f-f_0)\hat I(-f)\rangle$ \cite{GR2}. We have measured this quantity for $f\sim f_0$, in which case $\delta S_{I^2}$ measures how low frequency and high frequency currents become correlated by a high frequency excitation \cite{GR1}. We show on Fig. \ref{chi} the noise susceptibility $\chi$ that characterizes environmental contributions for $f=f_0=6$ GHz. It clearly differs from the low frequency result $\propto dS_{I^2}/dV$. From the knowledge of environmental effects in the quantum regime, we can extract the intrinsic quantum third moment of the current fluctuations that we have discussed before.

\section{Conclusion}

We have performed the first measurements of non-Gaussian noise in a conductor, both in the classical and quantum regimes. This work opens routes towards the study of higher cumulants of noise in complex devices to extract interesting physics (see, e.g. \cite{avalanches}) as well as it stresses the need for theoretical investigations to model measurements beyond average quantities. That a time-averaged measurement is predicted by the quantum and statistical average of the well suited operator is well established. What correlator (with the appropriate times and time-orderings) describes the measurement of a quantity at different times by a given setup is not yet solved \cite{Bednorz08, Bednorz09, Bayandin08, Bachmann09}.

\section*{Acknowledgements}

We are very grateful to J. Senzier for participating in the early stage of this work and to C. Wilson for providing us with a tunnel junction used for the first measurements. We acknowledge fruitful discussions with M. Aprili, C. Beenakker, A. Bednorz, W. Belzig, M. B\"uttiker, A. Clerk, M. Devoret, M. Kindermann, L. Levitov and Y. Nazarov.

\end{document}